\documentclass[aps,prl,twocolumn,groupedaddress]{revtex4}

\usepackage{graphicx}
\usepackage{amsmath}
\usepackage{gensymb}

\begin{document}


\title{Adiabaticity and localization in one-dimensional incommensurate lattices}


\author{E. E. Edwards, M. Beeler, Tao Hong, and S. L. Rolston}
\affiliation{Joint Quantum Institute and Department of Physics, University of Maryland, National Institute of Standards and Technology, College Park, MD 20742}


\date{\today}

\begin{abstract}
We experimentally investigate the role of localization on the adiabaticity of loading a  Bose-Einstein condensate into a one-dimensional optical potential comprised of a shallow primary lattice plus one or two perturbing lattice(s) of incommensurate period. We find that even a very weak perturbation causes dramatic changes in the momentum distribution and makes adiabatic loading of the combined lattice much more difficult than for a single period lattice.  We interpret our results using a band structure model and the one-dimensional Gross-Pitaevskii equation.    
\end{abstract}


\maketitle


Disorder plays an important role in many condensed matter systems,~\cite{Fisher,Anderson} with deep connections to quantum chaos~\cite{Grempel}, but can be difficult to systematically study due to the challenge of creating reproducible and quantifiable disorder.  The control available in ultra-cold atom systems~\cite{bloch} makes it an attractive platform to study disorder~\cite{Billy,Roati,Demarco,Schulte}.  To date much of the work adding disorder to ultra-cold atom systems has explored time-independent properties, but the long timescales associated with cold atoms allows investigation of dynamical properties as well (see \cite{Holthaus, DeSarlo, Lye, Billy} and ref. therein).  In this work we examine the ability of a quasi-disordered system to adiabatically follow changes in the Hamiltonian.  The presence of disorder produces a complicated eigenvalue spectrum, which greatly affects the adiabaticity criteria.  The physics  of localization phenomena also has a significant impact on time-dependent processes, such as adiabaticity.  Small perturbations to the Hamiltonian can cause large changes to the ground state wavefunction over large length scales, making it difficult for the system to adiabatically follow changes.  One recent theoretical study shows that adiabaticity in gapless systems is non-trivial, particularly in lower dimensions~\cite{Polk}.  Here we show that even in  a gapped system such as ours, adiabaticity is complicated by the presence of disorder. 

We study adiabaticity in a quasi-disordered system by adding one or two weak incommensurate lattices to a one-dimensional optical lattice loaded with a Bose-Einstein condensate (BEC).  Localization occurs in both  disordered and strictly incommensurate potentials~\cite{Roth, sanchez2, damski, Schulte} although with distinct differences, which tend to disappear in finite-sized systems such as ours.  We observe a complex momentum distribution of the atoms due to the presence of weak perturbing lattices following a ramped loading process that would be nearly adiabatic for a single lattice.  We gain insight into the distributions from single-particle band structure, and observe that the effects of the perturbations disappear as interactions increase, as they suppress the long wavelength density modulation of the wavefunction.

We form a BEC of $\sim$$10^4$ $^{87}$Rb atoms  in the state $|F=2,m_{f}=2\rangle$ in a magnetic trap with $\omega_{x}$ $\approx \omega_{z}$ $\approx2\pi$$\times$$410$ Hz and $\omega_{y}$ $\approx2\pi$$\times$$120$ Hz. To reduce the effects of interactions, the trap is subsequently weakened giving final frequencies $\omega_{x}$ $\approx2\pi$$\times$$ 40$ Hz,  $\omega_{y}$ $\approx2\pi$$\times$$ 20$ Hz, and  $\omega_{z}$ $\approx2\pi$$\times$$ 30$ Hz.  We load the BEC into a 1D optical potential, created by the addition of a primary and perturbing lattice(s).  The total potential is 
\begin {align}
V_{tot}=\frac{M}{2}(\omega_{x}^2x^2+\omega_{y}^2y^2+\omega_{z}^2z^2)+V_{1}\sin^2 (k_{1}z)\nonumber \\
+V_{2}\sin^2 (k_{2}z)+V_{3}\sin^2 (k_{3}z),
\end{align} 
 where $M$ is the atomic mass, $k_1=2\pi/ \lambda$, and $\lambda=796.6$ nm.  The ratios $k_{2}/k_{1}= 0.806\pm0.002$ and $k_3$/$k_1=0.919\pm0.004$ are extracted from images of atomic diffraction.  For the bulk of the experiments described here, the lattice depths are $V_1=4.6\pm0.3$ $E_{rec}$ and $V_{2,3}/V_{1}=0.059\pm0.003$, respectively, where $E_{rec} = \hbar^2 k_{1}^2/2M$.  To prevent interference between the lattices, we detune the beams several MHz from each other with acousto-optic modulators, so that the coupling terms between the beams oscillate rapidly compared to the atomic motion.  The lattices are created by reflection off an in-vacuum gold mirror situated approximately 2 mm from the magnetic trap center.  Each beam is reflected at a different angle ($\theta_{1}=180$$\degree$, $\theta_{2}=143\degree$, and $\theta_{3}=155\degree$) to yield a 1D pseudo-disordered potential~\cite{Guidoni}.  Because the gold mirror surface that defines the standing waves nodes is the same for all three lattices, they are phase-locked together.  We load the atoms into the combined lattices by ramping up the intensities linearly, keeping a fixed ratio between $V_{2,3}$ and $V_{1}$.  A ramp time of 1 ms is chosen for most experiments  and is sufficiently long to ensure loading a magnetically trapped BEC into the lowest band of the primary lattice.  In order to remain adiabatic, the excited state population, $|a_{n,q}(t)|^2$, must stay $\ll1$, where $n$ is the band index, and $q$ is the quasimomentum.  One can calculate a corresponding time scale using
\begin {equation}\dot{a}_{n,q}(t)=\frac{a_{1,q}\hbar}{(E_{n,q}-E_{1,q})}\big<n,q|\frac{\partial H}{\partial t}|1,q\big> 
e^{i\int_{0}^t{\frac{E_{n,q}-E_{1,q}}{\hbar}}dt}
\end{equation}
which, for loading a $^{87}$Rb BEC into a single lattice at $k_1\approx2\pi/\lambda$, is satisfied for times much greater than 5 $\mu$s~\cite{Schiff,P.S.Julienne,Peil}.  This short time scale is because  the nearest excited band with allowed transitions is approximately 4 $\hbar^2 k^2/2M$ separated from the ground band  with initial $q=0$, even for an arbitrarily weak lattice with reciprocal lattice vector of $2k$.  As the lattice depth increases, the bands continue to separate.  Adiabaticity becomes difficult for q near the band edge, where the initial energy gap vanishes.  For our experiments, the initial momentum range is $\pm 0.15$ $k_1$.   Additionally, the single particle band picture begins to break down as the BEC becomes strongly interacting. 

To assess our ability to adiabatically load a lattice, we perform two types of experiments.  In all experiments,  we  absorption image the cloud after 22 ms of time-of-flight.   Following the ramped loading and a variable hold time, we abruptly turn off both the lattice and the magnetic trapping potentials.  The image yields the momentum distribution of the atoms trapped in the lattice (interactions during time-of-flight are negligible).  In the second method, we slowly ramp down the lattice and then release from the magnetic trap. If the process is fully adiabatic, the cloud should return to the momentum distribution of the the original BEC.  For comparison, we present absorption images after the sudden turn-off, with and without the perturbative lattice(s) in Fig. 1.  The existence of the weak perturbing lattice (Fig.1(c-e)) markedly modifies the momentum distribution with the appearance of new peaks, even though the perturbing lattice is so weak that it does not produce any observable diffraction alone (Fig.1(b)).  In Ref.~\cite{Roth} it was suggested that an indication of localization is the appearance of additional momentum peaks in the matter-wave interference pattern.  
\begin{figure}
\scalebox{0.45}{\includegraphics{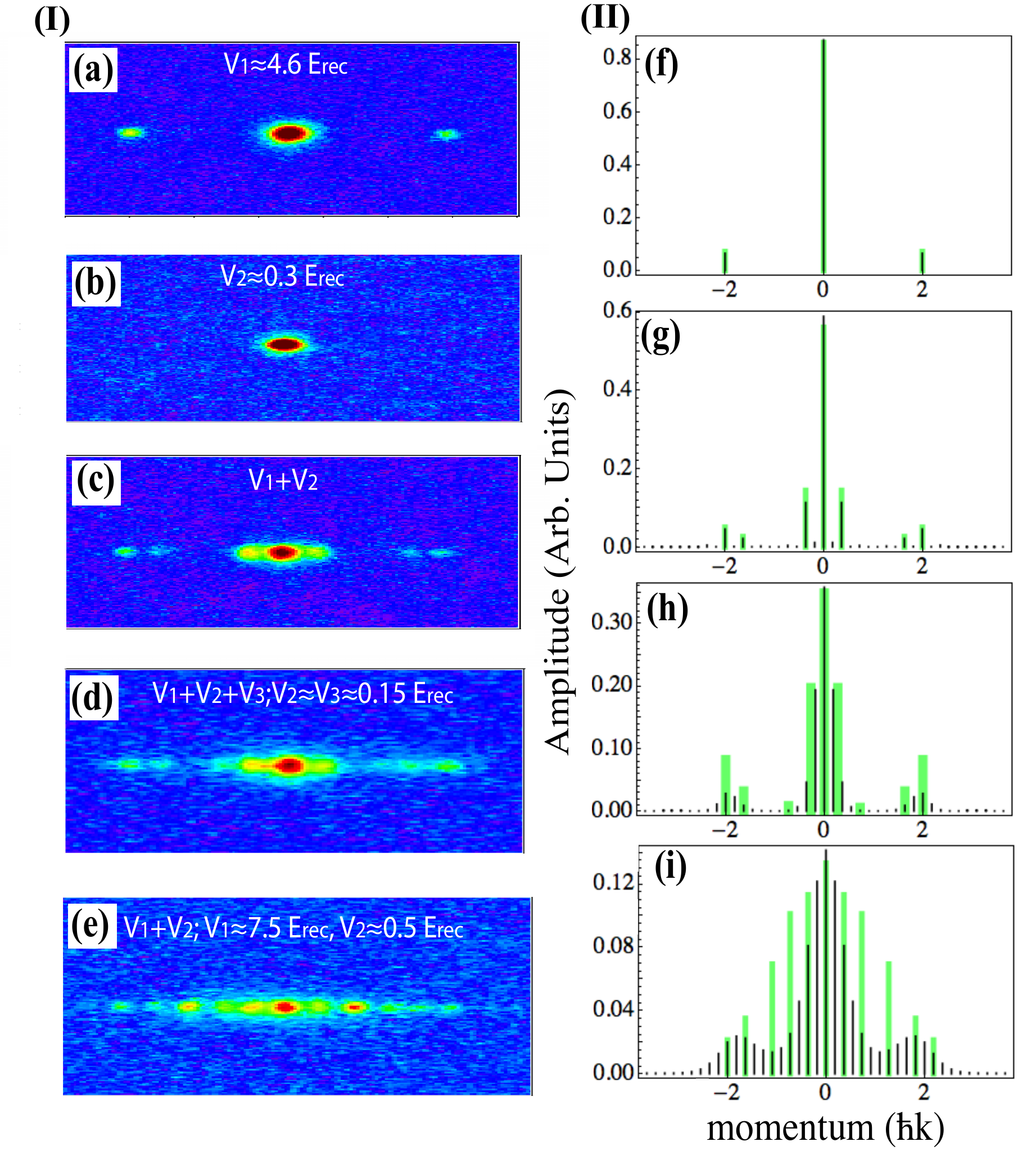}}
\caption{\label{fig:1}(I) Absorption images of a BEC loaded into different lattices for $\mu_{BEC}\approx100$ Hz after 20 ms TOF:  (a)$V_1$ = $4.6$ $E_{rec}$ and (b) $V_2$ = $0.3$ $E_{rec}$;(c) $V_1+V_2$; (d) $V_1+V_2+V_3$, with $V_2$ = $V_3$ = $0.15$ $E_{rec}$  ; (e) $7.5$ $E_{rec}$ and $0.5$ $E_{rec}$ lattice.  We turn on lattices using a 1 ms linear ramp followed by a 1 ms hold and abrupt turn-off.  (II) Using the data from the absorption images ((I)-(a,c-e)) we extract the relative population amplitudes and overlay them with a band structure calculation (no interactions) of the ground state momentum distribution. 
}
\end{figure}

In order to understand why such a small perturbation has dramatic effects on the momentum structure and excited fraction we use single-particle band structure. Even if the perturbing lattice is strictly incommensurate with the primary lattice ($k_{2}/k_{1}=\alpha$, an irrational number), we can approximate $\alpha$ as a ratio of two large integers, $f/g$.  In a finite system such as ours where the BEC occupies $\approx70$ sites of the primary lattice and does not extend over many periods of the beat frequency lattice (created by the combination of the lattice potentials) we expect this approach to yield reasonable predictions~\cite{Diener}.

Assuming no excitations to higher bands, we predict the momentum distribution for each lattice configuration using a band structure calculation with $f/g=9/11$ and $f^{\prime}/g=10/11$,  approximations to the experimental ratios.  We present the results of this calculation over-layed with population amplitudes (green bars) extracted from fits of our data in Fig. 1.   In the case of three lattices, experimental resolution limits our ability to resolve the structure, and is represented by wider bars. The combination of the lattices creates a complex momentum structure, and the band structure calculation is in good qualitative agreement with our measurements.  However, since the band-structure calculation assumes no excitations and a 1 ms ramp causes depletion of the ground band, the predicted populations do not quantitatively agree with the data for multiple lattices.  This mismatch increases as the depths of both the primary and perturbing lattice(s) are increased.  For deep lattices (Fig. 1(i)), depending on the choice of $f/g$ , band structure  predicts peaks spaced at $k_{1}/g$, closer than the beat frequency, yet the envelope does not change. We find these disappear much more rapidly than the beat frequency peaks as interactions increase (see below).   

We also examine the changing wavefunction and the corresponding band-structure in Fig. 2.  Using the $k-$vector of the primary lattice to determine the size of the first Brillioun zone, we plot the first two bands for the case of one and two lattices.  The dominant effect of the perturbing lattice is to open small gaps in the ground band, as well as to slightly flatten the band. Starting at $q=0$ and traveling along the band, the first gap corresponds to quasimomentum of $(k_{1}-k_{2})$. If instead one chooses the first Brillioun zone to span $\pm(k_1-k_2)$, there exists a new band that is much lower in energy than the first excited band in the single lattice case, which determines a new energy scale relevant to adiabaticity.  This new energy scale is also responsible for localizing the wavefunction at the spatial period corresponding to the beat frequency (Fig. 2(c) compared to Fig. 2(b)).  In order to satisfy the adiabaticity condition (Eq. 2) for two lattices with our parameters ($V_2\approx0.059$ $V_1, k_1/k_2\approx0.8$), the loading ramp must be much longer than $4$ ms, a thousand times longer than for a single lattice.  This is shown in Fig. 2(e), where we calculate the depletion of population in the ground band for different ramp rates as a function of primary lattice depth for the case of three lattices.  The inset of Fig. 2(e) depicts an expanded view of ramp times yielding less than 5 percent excitation.  Figure 2(f) shows the depletion of the ground state as a function of lattice depth for a ramp time of approximately 2.5 ms for the case of one, two and three lattices.  

Since it is often the case experimentally that we start with a BEC with non-zero $q$ (due to residual motion in the magnetic trap), we also show curves for loading a BEC with $q$ = 0.1  $k_1$ into a potential comprised of two and three lattices.  This value of $q$ was chosen to lie within the new, smaller first Brillioun zone.  These curves show that the effect of additional perturbing lattices is the dominant effect on adiabaticity criteria.  For $q$ = 0.1 $k_1$, the adiabaticity time scale approaches $10$ ms, compared to virtually no change for a single lattice with initial $q$ of 0.1 $k_1$. We experimentally observe that by ramping up the potential and then ramping down with additional lattices,  longer ramp times (5 ms compared to 250 $\mu$s) are qualitatively better (less excitations), but due to interactions, we are never able to be fully adiabatic.  Although excitation to higher bands may also produce new momentum peaks, for the data presented in Fig. 1(c), we predict that between 1 and 5 percent of the population is excited depending on initial quasimomentum. Thus most of the observed changes in the momentum distribution can be ascribed to modification of the ground band wavefunction, i.e. localization.

The effects seen here are a result of the incommensurability of the lattices.  For experiments using commensurate lattices ~\cite{Peil} where two lattices had a carefully chosen ratio of 3:1, the criteria for adiabaticity with respect to the band structure is not significantly altered.  In this case the higher order momentum peaks due to the perturbing lattice overlap the beat frequency peaks, which results in the wavefunction having modulation only at that spatial frequency. This difference can be seen by calculating the ground state wavefunction for the two cases.  For approximately incommensurate lattices, the perturbation has a strong effect, with small spatial frequencies appearing (Fig. 2(c)).   In order to further quantify this dramatic change, we plot the population of the $p=0$ peak, $p_{0}$, as a function of lattice depth for a single lattice, and for a primary lattice with a perturbative lattice (with $V_2=0.059$ $V_1$).   The black points are experimental measurements of the peak amplitude, which show a dramatic reduction of $p_{0}$ when the perturbing lattice is added.  We show two band structure calculations for $f/g$=$4/5$ (dashed) and $9/11$ (solid), both falling within the error bars of our experimental ratios.  The significant difference between these cases is due to population growth in sub-beat frequency spaced momentum peaks predicted for the ratio $9/11$.  The error bars represent statistical uncertainties in populations and lattice depths (due to fluctuations in lattice beam intensity).  We expect that because the turn-on of two lattices is not fully adiabatic there will be additional depletion of the central peak due to excitations into higher bands, as observed. 
\begin{figure}
\scalebox{0.4}{\includegraphics{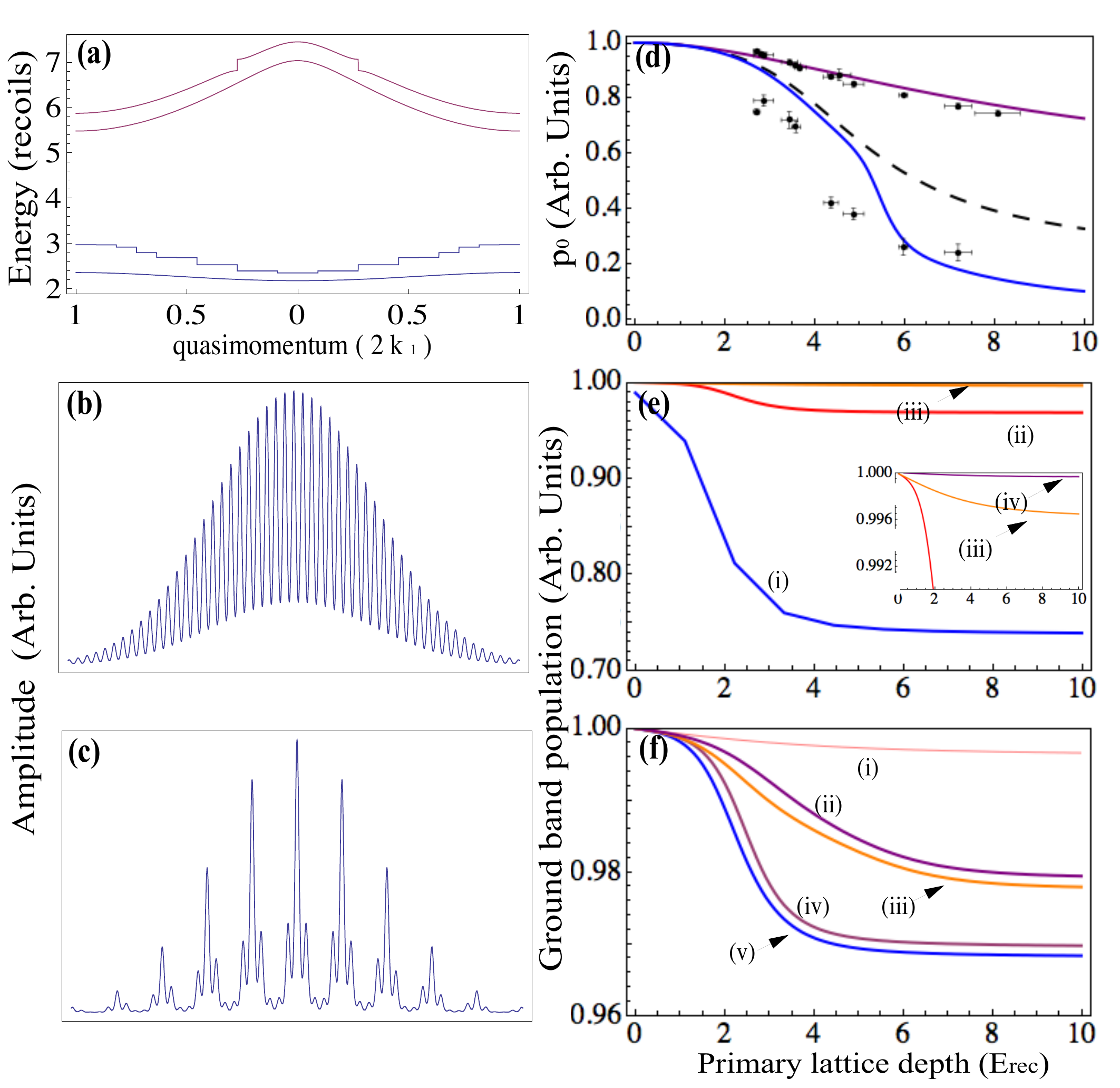}}
\caption{\label{fig:2} (a) Calculated first two bands for a single lattice and for two incommensurate lattices.  Note the gaps which form in the ground band. (b-c) Plot of an example wavefunction multiplied by a Gaussian envelope (b) without and (c) with a perturbing lattice and (d) the probability amplitude, $p_{0}$, as a function of lattice depth for $p=0$ peak for the case of a single lattice and a combined potential. The changes in slope correspond to the depths at which the atoms become localized at different spatial frequencies.  (e)  Calculation of ground band population for differing ramp time ((i) 2.5 $\mu s$, (ii) 250 $\mu s$, (iii) 2.5 ms) for case of three lattices; (iv) for loading a single lattice with a ramp time of 2.5 ms.   Inset in (e) is a zoomed-in view for comparison of (iii) and (iv).  (f) Calculated ground state population for loading a (i) single lattice, (ii) two lattices, (iii) two lattices starting with $q=0.1$ and three lattices starting with (iv) $q=0$ and (v) $q=0.1$ for a fixed ramp time of 2.5 ms.}
\end{figure}

The production of low spatial frequency, long-wavelength components in the wavefunction with the application of weak incommensurate lattices yields insight into the connection between disorder, localization, and incommensurate systems.  In the canonical view of localization, destructive interference due to reflections over long distances leads to the localization of the wavefunction~\cite{Anderson,Sanchez}.  Here we see that a single incommensurate perturber yields a complex long wavelength structure.  If we add a second perturber, getting closer to true disorder, we see even more momentum components.  Experimentally, due to finite resolution, this appears as a broadening of the distribution (Fig. 1(d)).  Comparing three lattices to two, we observe a larger spread in the central feature, as well as a corresponding decrease in optical depth.  This indicates that there are unresolved peaks beneath the overall envelope.  If something competes with these long range interferences, such as atom-atom interactions, we can expect localization phenomena to be drastically modified~\cite{Lye}.  Indeed, until recently~\cite{Roati, Billy} Anderson localization had not been seen in cold atom systems due to the effect of atom-atom interactions.  

To study the effects of interactions in our system, we perform experiments in three different magnetic traps of differing frequencies with chemical potentials of $\mu_{BEC}/h$$\approx$$(2500,400,100)$ Hz.  We observe (Fig. 3) that as the overall confinement is increased, both the structure in the interference pattern such as the $k_{1}-k_2$ momentum peaks disappears and the overall size of the cloud, after TOF, increases.

We compare the results of the non-interacting band-structure model to simulations of our system using the Gross-Pitaevskii equation (GPE).  We reduce the 3D equation to 1D and solve using the split-operator method in combination with imaginary time evolution~\cite{Weizhu}.  The simulations using the GPE predict that in the moderately tight trap ($\mu_{BEC}/h=400$ Hz), the $k_1-k_2$ (beat frequency) peaks slightly persist, however they disappear in the tightest trap.  Fig. 3 shows absorption images with results of GP simulation in three different traps. In our system, we cannot decouple the size of the sample from interactions, which both contribute to the amount of spatial localization.  This can be done if one utilizes a Feshbach resonance, as in~\cite{Roati}, to vary the strength of the scattering length.  We calculate the effect of arbitrarily increasing the interactions in our weakest trap (where the structure is most apparent) and see that the sub-beat frequency peaks vanish rapidly, followed by the $k_1-k_2$ peaks.  This can be understood in that interactions drive transitions between states of different quasimomentum, washing out the discrete low momentum features, and destroying the long range spatial periodicity of the wavefunction. 
\begin{figure}
\scalebox{0.4}{\includegraphics{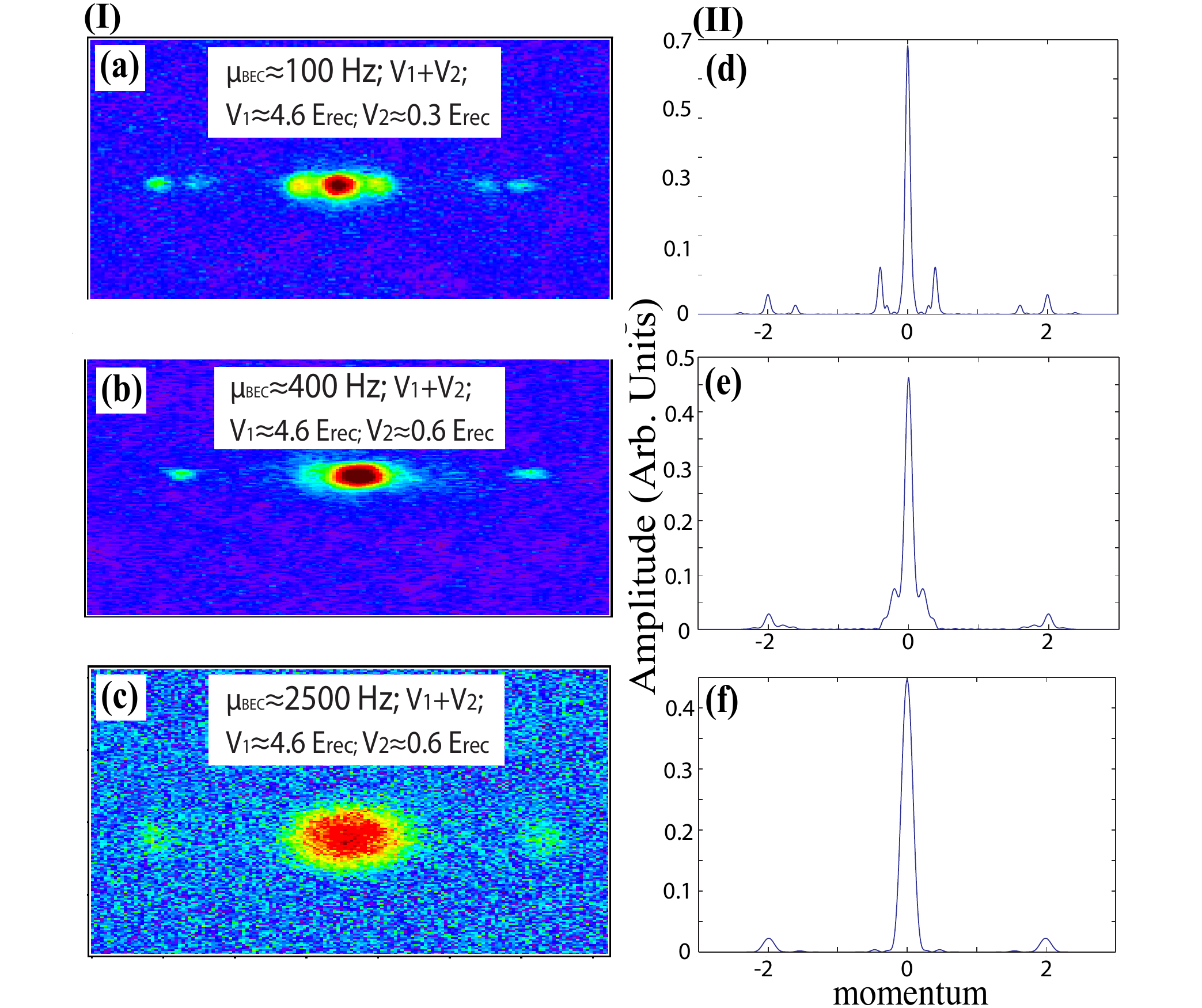}}
\caption{\label{fig:4} (I) (a-c) Absorption images in a traps with $\mu_{BEC}\approx(100, 400, 2500)$ Hz with a primary plus one perturbing lattices. (II) (d-f) Results of GP simulation for (a-c) showing the effects of interactions.}
\end{figure}

In conclusion, we have presented data indicating that small perturbations to a one-dimensional lattice system in the form of quasi-disorder, while leading to  localization of the wavefunction, also drastically changes the dynamics of the system.  We present a theoretical treatment which suggests that although we are only slightly modifying the energy of the system, the large alterations of the wavefunction demand time scales for adiabaticity that are orders of magnitude longer than for loading a single lattice.  Because of the sensitivity of the adiabaticity criteria in the presence of perturbations, one should be careful when studying disordered lattice systems to identify and characterize any forms of disorder present in the potential, intentional or otherwise.  This work shows that disorder can have a strong influence on dynamics,  and that the long timescales of cold-atom optical lattice systems makes them ideal for further explorations.

We acknowledge J. Robinson, I. B. Spielman, and J.V. Porto for technical contributions and useful discussions.  This work was supported by ARO, NSF, and NASA.

\end{document}